\input harvmac

\def\eql{~=~}
\def\seql{\! = \!}
\def\al{\alpha}
\def\ga{\gamma}

\def\cN{{\cal N}}

\def\ie{{\it i.e.}}
\def\bfone{\relax{\rm 1\kern-.35em 1}}
\def\IC{\relax\,\hbox{$\inbar\kern-.3em{\rm C}$}}
\def\ID{\relax{\rm I\kern-.18em D}}
\def\IF{\relax{\rm I\kern-.18em F}}
\def\IH{\relax{\rm I\kern-.18em H}}
\def\II{\relax{\rm I\kern-.17em I}}
\def\IN{\relax{\rm I\kern-.18em N}}
\def\IP{\relax{\rm I\kern-.18em P}}
\def\IQ{\relax\,\hbox{$\inbar\kern-.3em{\rm Q}$}}
\def\us#1{\underline{#1}}
\def\IR{\relax{\rm I\kern-.18em R}}
\font\cmss=cmss10 \font\cmsss=cmss10 at 7pt
\def\ZZ{\relax\ifmmode\mathchoice
{\hbox{\cmss Z\kern-.4em Z}}{\hbox{\cmss Z\kern-.4em Z}}
{\lower.9pt\hbox{\cmsss Z\kern-.4em Z}}
{\lower1.2pt\hbox{\cmsss Z\kern-.4em Z}}\else{\cmss Z\kern-.4em
Z}\fi}
%
%
\def\nihil#1{{\it #1}}
\def\eprt#1{{\tt #1}}
\def\nup#1({Nucl.\ Phys.\ $\us {B#1}$\ (}
\def\plt#1({Phys.\ Lett.\ $\us  {#1B}$\ (}
\def\cmp#1({Comm.\ Math.\ Phys.\ $\us  {#1}$\ (}
\def\prp#1({Phys.\ Rep.\ $\us  {#1}$\ (}
\def\prl#1({Phys.\ Rev.\ Lett.\ $\us  {#1}$\ (}
\def\prv#1({Phys.\ Rev.\ $\us  {#1}$\ (}
\def\mpl#1({Mod.\ Phys.\ Let.\ $\us  {A#1}$\ (}
\def\ijmp#1({Int.\ J.\ Mod.\ Phys.\ $\us{A#1}$\ (}
\def\jag#1({Jour.\ Alg.\ Geom.\ $\us {#1}$\ (}


\lref\KPW{A.\ Khavaev, K.\ Pilch and N.P.\ Warner, \nihil{New Vacua of
Gauged  ${\cal N}=8$ Supergravity in Five Dimensions},
\eprt{hep-th/9812038}.}
\lref\GRW{M.\ G\"unaydin, L.J.\ Romans and N.P.\ Warner,
\nihil{Gauged $N=8$ Supergravity in Five Dimensions,}
Phys.~Lett.~{\bf 154B} (1985) 268; \nihil{Compact and Non-Compact
Gauged Supergravity Theories in Five Dimensions,}
\nup{272} (1986) 598.}
\lref\PPvN{M.~Pernici, K.~Pilch and P. van Nieuwenhuizen,
\nihil{Gauged $N=8$, $D = 5$ Supergravity,} \nup{259} (1985) 460.}
\lref\LJR{L.J.\ Romans, \nihil{New Compactifications of Chiral $N=2$,
$d=10$ Supergravity,}  \plt{153} (1985) 392.}
\lref\RLMS{R.~G. Leigh and M.~J. Strassler, \nihil{Exactly Marginal
Operators and
Duality in Four-Dimensional $N=1$ Supersymmetric Gauge Theory,}
\nup{447} (1995) 95; \eprt{hep-th/9503121}}
\lref\FGPWa{D.~Z. Freedman, S.~S. Gubser, K.~Pilch, and N.~P. Warner,
\nihil{Renormalization Group Flows from Holography---Supersymmetry and a
c-Theorem,} CERN-TH-99-86, \eprt{hep-th/9904017} }
\lref\JSIIB{J.H.\ Schwarz, \nihil{Covariant Field Equations of
Chiral $N=2$, $D=10$ Supergravity,} CALT-68-1016,
\nup{226} (1983) 269.}
\lref\PvNW{P.~van~Nieuwenhuizen and N.P. Warner, \nihil{New
Compactifications of Ten-Dimensional and Eleven-Dimensional
Supergravity on Manifolds which are not Direct Products} \cmp{99}
(1985) 141.}
\lref\CLP{M.~Cveti\v{c}, H.~L\"u and C.N.~Pope, \nihil{Geometry of
the
Embedding of Scalar Manifolds in $D=11$ and $D=10$,  }
\eprt{hep-th/0002099}.}
\lref\MetAns{B.\ de Wit and H.\ Nicolai, \nihil{On the Relation
Between $d=4$ and $d=11$ Supergravity,} Nucl.~Phys.~{\bf B243}
(1984) 91; \hfil \break
B.\ de Wit, H.\ Nicolai and N.P.\ Warner,
\nihil{The Embedding of Gauged $N=8$ Supergravity into $d=11$
Supergravity,}
Nucl.~Phys.~{\bf B255} (1985) 29.}
\lref\adscftrev{
O.~Aharony, S.~S.~Gubser, J.~Maldacena, H.~Ooguri and Y.~Oz,
\nihil{Large N Field Theories, String Theory and Gravity,}
\eprt{hep-th/9905111}. }
\lref\DiLithium{B.\ de Wit, H.\ Nicolai,  \nihil{A New $SO(7)$
Invariant Solution of $d = 11$ Supergravity,}
Phys.~Lett.~{\bf 148B} (1984) 60.}
\lref\NWstrings{N.P.\ Warner, \nihil{Renormalization Group Flows from
Five-Dimensional Supergravity,} talk presented at Strings `99, Potsdam,
Germany, 19--25 Jul 1999; \eprt{hep-th/9911240}}
\lref\PosEn{L.\ F.\ Abbott and S.\ Deser,  \nihil{Stabilty of Gravity
with
a Cosmological Constant,}  Nucl.~Phys.~{\bf B195} (1982) 76;
\hfil\break
G.W.\ Gibbons, C.M.\ Hull and N.\ P.\ Warner,  \nihil{The Stability
of Gauged Supergravity}  Nucl.~Phys.~{\bf B218} (1983) 173;
\hfil\break
W.\ Boucher, \nihil{Positive Energy without Supersymmetry,}
Nucl.~Phys.~{\bf B242} (1984) 282; \hfil \break
L.\ Mezincescu and P.K.\ Townsend, \nihil{Positive Energy and the
Scalar
Potential in Higher Dimensional (Super)Gravity Theories,}
Phys.~Lett.~{\bf 148B} (1984) 55; \hfil \break
L.\ Mezincescu and P.K.\ Townsend, \nihil{Stability  at a Local
Maximum
in Higher Dimensional Anti-de Sitter Space and Applications to
Supergravity,}
Ann.~Phys.~{\bf 160} (1985) 406.}
%
\lref\KLM{A.~Karch, D.~Lust and A.~Miemiec,
\nihil{New N = 1 Superconformal Field Theories and their Supergravity  
Description},
Phys.\ Lett.\  {\bf B454} (1999) 265, \eprt{hep-th/9901041}.}
\lref\StGu{S. Gubser, private communication.}
\lref\MaSt{M. Strassler, private communication.}

\Title{
\vbox{
\hbox{CITUSC/00-012}
\hbox{USC-00/01}
\hbox{\tt hep-th/0002192}
}}
{\vbox{\vskip -1.0cm
\centerline{\hbox{A New Supersymmetric Compactification of}}
\vskip 8 pt
\centerline{\hbox{Chiral IIB Supergravity }}}}
\vskip -.3cm
\centerline{Krzysztof Pilch and Nicholas P.\ Warner }
\medskip
\centerline{{\it Department of Physics and Astronomy}}
\centerline{{\it and}}
\centerline{{\it CIT-USC Center for Theoretical Physics}}
\centerline{{\it University of Southern California}}
\centerline{{\it Los Angeles, CA 90089-0484, USA}}

\bigskip
\bigskip
We present a new compactification of chiral, $\cN \seql 2$
ten-dimensional supergravity down to five dimensions and show that it
corresponds to the $\cN \seql 2$ supersymmetric critical point of
five-dimensional, $\cN \seql 8$ gauged supergravity found in \KPW.
This solution presented here is of particular significance because it
involves non-zero tensor gauge fields and, via the AdS/CFT
correspondence, is dual to the non-trivial $\cN \seql 1$
supersymmetric fixed point of $\cN \seql 4$ Yang-Mills theory.

\vskip .3in
\Date{\sl {February, 2000}}

%
\parskip=4pt plus 15pt minus 1pt
\baselineskip=15pt plus 2pt minus 1pt

\newsec{Introduction}

The generalizations of the AdS/CFT correspondence (see, e.g.,
\adscftrev\ and the references therein) to theories with a scale has
proven a rather successful enterprise over the last year.  In
particular, much work has been done on using supergravity to describe
renormalization group flows of large $N$ field theories.  One of the
basic approaches to this has been to take a conformal field theory for
which the correspondence has been well established and then perturb it
by one or more relevant operators and use supergravity to follow the
flow to the infra-red.  The first supersymmetric flow to be described
in terms of supergravity was discussed in \FGPWa: The starting point
of the flow was $\cN = 4$ Yang-Mills theory, while the end-point was
an $\cN=1$ superconformal fixed point  identified in field theory
in \refs{\RLMS,\StGu,\KLM,\FGPWa}.

Five-dimensional, $\cN=8$ gauged supergravity is an invaluable tool in
studying flows of the $\cN=4$ Yang-Mills theory.  The former field
theory arises in the $S^5$ compactification of IIB supergravity as the
field theory of the five-dimensional, $\cN=8$ graviton supermultiplet.
This is dual to the energy-momentum tensor supermultiplet of $\cN=4$
Yang-Mills, and in particular the supergravity scalars are dual to
gauge invariant, bilinear operators of the Yang-Mills theory. Thus
supergravity scalars can be used to describe mass terms and vevs as
well as the Yang-Mills coupling.  The important point is that the
five-dimensional, $\cN=8$ gauged supergravity is, almost certainly a
consistent truncation of the ten-dimensional, IIB theory, and
therefore the equations of motion of the five-dimensional theory are
embedded in those of the ten-dimensional theory to the extent that a
solution of the former implies a solution of the latter.  This means
that the complexities of the ten-dimensional theory (on this special
subsector) can be simplified by working entirely with the
five-dimensional theory.

This was the approach taken in \KPW, where new critical points of the
five-dimensional scalar potential were given.  According to consistent
truncation, these critical points must correspond to solutions of the
IIB theory, and indeed this has been shown for the simpler critical
points \refs{\PvNW,\LJR}.  From the AdS/CFT correspondence one should
expect these critical points to correspond to new phases of the
Yang-Mills theory.  There is, however, one caveat: vacua that are
unstable in supergravity exhibit apparently non-unitary behaviour in
the corresponding Yang-Mills phases.  At present the meaning of
critical points is less than clear, but as vacua they are
pathological, and so should be discarded.  On the other hand,
supersymmetric vacua in supergravity are known to be completely
semi-classically stable \PosEn\ \foot{By an interesting
``coincidence''(?)  the only known perturbatively stable critical
points are the supersymmetric ones.}.
It was argued in \refs{\KPW,\NWstrings} that supersymmetric
critical points should always represent new phases of the Yang-Mills
theory {\it even at finite N}.  This is because supersymmetric ground
states of IIB supergravity should provide ground states for the IIB
string, and hence the Yang-Mills phase should survive at finite N.  By
extrapolation, the non-supersymmetric, unstable critical points cannot
be good string vacua, and so are most likely an example of ``large N
pathology.''

It was for the foregoing reasons that the new $\cN=2$ supersymmetric
critical point in \KPW\ was particularly significant.  Moreover the
subsequent analysis of that critical point, and the flow to it in
supergravity matched perfectly with corresponding results in field
theory providing a highly non-trivial test of the extension of the
AdS/CFT correspondence to theories with a scale.

Our purpose in this paper is to return to the supersymmetric critical
point in \KPW\ and construct the exact corresponding ten-dimensional
solution.  In five dimensions this solution corresponded to turning on
two scalar fields, but in ten-dimensions the background is rather more
complex.  The metric is deformed from the round $S^5$, and there are
warp factors, but the really new feature is that the non-zero fermion
masses in the Yang-Mills theory mean that there must be non-zero
$B_{\mu \nu}$ fields in ten-dimensions.  There are few, if any, such
backgrounds known that are also supersymmetric.  Thus it is
intrinsically interesting to exhibit such a ten-dimensional solution.
Moreover, it is even more significant in that this solution represents
the supergravity (and string theory) dual of the Leigh-Strassler fixed
point.  Preliminary work towards finding the ten-dimensional solution
was done in \CLP, where the metric was computed using the formula
given in \KPW, and a linearized form for the $B$-field was suggested.

Our analysis will proceed as follows: In section 2 we first obtain the
ten-dimensional metric using the result in \KPW, and we
give the Ricci tensor for this metric.
Then, in section 3 we use the symmetries of the five-dimensional
solution to arrive at an Ansatz for the tensor gauge fields.
We then solve the ten-dimensional equations and exhibit the
solution.  In section 4 we confirm that our solution does
indeed have the proper amount of supersymmetry, and give
explicit formulae for the generators.

\newsec{The scalars and the metric}

We first, briefly review the structure of the five-dimensional
solution, paying particular attention to its symmetries as they will
be important later.

\subsec{The five-dimensional solution and its symmetries}

In terms of Yang-Mills theory, the solution involves giving a mass to
s single $\cN=1$ hypermultiplet.  That is, one perturbs the
Hamiltonian by the fermion bilinear $Tr(\lambda^4 \lambda^4) $, and by
the scalar field counterpart: $Tr(X^5 X^5 + X^6 X^6)$.  We will denote
the corresponding supergravity fields by $\chi$ and $\alpha$,
respectively.%
\foot{In \FGPWa\ these were denoted by $\varphi_1$
and $\alpha$.}  The first of these fields breaks the $SO(6)$
$R$-symmetry to $SU(3)$, while the second field breaks the $SO(6)$
down to $SO(4) \times SO(2)$.  The two together break the $SO(6)$
to the group $SU(2) \times U(1)$, where the $U(1)$ is the
$R$-symmetry at the new $\cN=1$ Yang-Mills fixed point.

To see this more explicitly, think of $SO(6)$ as $SU(4)$ on the
fundamental, then the $SU(3)$ is the upper-left $3 \times 3$ block,
while $SO(4) \times SO(2) \equiv SU(2) \times SU(2) \times U(1)$ acts
as two $2 \times 2$ blocks, with the $U(1)$ as ${\rm diag}(e^{i
\phi},e^{i \phi},e^{-i \phi},e^{-i \phi})$.  The $SU(2) \times U(1)$
invariance is the common subgroup consisting of the upper-left $SU(2)$
and the $U(1)$: ${\rm diag}(e^{i \phi},e^{i \phi},e^{-2 i \phi},1)$.

If one takes into account the $SL(2,\IR)$ symmetry of the supergravity
then there is a further $U(1)$ symmetry.  The supergravity scalar
$\alpha$ is $SL(2,\IR)$ invariant, but $\chi$ is part of an
$SL(2,\IR)$ doublet, and may be taken to have charge $+1$ under
$U(1)\subset SL(2,\IR)$.  This can be cancelled by the action of the
$U(1)$ factor in $SU(3) \times U(1) \subset SU(4)$.

On this two parameter subspace the supergravity potential
takes the form \KPW:
\eqn\potredcd{
V ~=~ -{g^2 \over 4 \rho^{4}} ~(1 - \sinh^4(\chi) ) ~-~
{g^2 \over 2 } ~\rho^{2}~ \cosh^2(\chi)  ~+~
{g^2 \over 32} ~ \rho^{8}~  \sinh^2(2 \chi)     \ ,}
and the superpotential is:
\eqn\Wreduced{W~=~ {1 \over 4 \rho^2}~ \Big[\cosh(2 \chi)~
( \rho^{6}~-~ 2)~ - ( 3\rho^{6} ~+~ 2 ) \Big] \ .}
where $\rho = e^\alpha$.  The potential and superpotential
are related via:

\eqn\VfromW{V ~=~ {g^2 \over 8}~ \Big| {\del W
\over \del \chi} \Big|^2~ + {g^2 \over 48}~ \Big| {\del W
\over \del \alpha} \Big|^2 ~-~ {g^2 \over 3}~\big|W \big|^2 \ .}
The $\cN=2$ supersymmetric critical point occurs at:
\eqn\critpt{ \chi\eql \log(3)/2\,,\qquad \alpha\eql \log(2)/6\,.}
Let $\Lambda$ and $\Lambda_0$ be the cosmological constants
at the new critical point, and at the maximally supersymmetric point
($\alpha =\chi =0$) respectively, then:
\eqn\cosmorat{ {\Lambda \over \Lambda_0} \eql {V\over V_0} \eql
{4 \, 2^{4/3} \over 9}  \,.}

\subsec{The metric in ten-dimensions}

The ten-dimensional space-time we seek is a ``warped'' product
$AdS_5\times M_5$ with the metric of the form:\foot{We use the
$(+,-,\ldots,-)$ convention for the metric.}
\eqn\metrten{ ds_{10}^2\eql \Omega^2 \,ds^2_{{\rm AdS}}-ds_5^2\, ,}
where $ds^2_{{\rm AdS}}$ is the metric on $AdS_5$, $ds_5^2$ is the
metric on the internal manifold, $M_5$, which is topologically a sphere.

The function, $\Omega$, is the warp factor and it depends on the
coordinates
of the internal manifold.  Following \FGPWa, we take a coordinate
system in which the AdS metric takes the form:
\eqn\adsmetr{
ds^2_{{\rm AdS}}\eql e^{2 A(r)}(dx_\mu dx^\mu)-dr^2\,,\qquad A(r)\eql
{r\over L}\,, }

Consistent truncation directly determines the internal metric of
\metrten\ in terms of the scalar fields.  The formula was obtained in
\KPW, and its derivation is exactly parallel to the corresponding
four-dimensional result \MetAns.  The formula expresses the deformed
metric in terms of bilinears in the Killing vectors, and thus it is
simplest to start by taking coordinates in which these Killing vectors
are as simple as possible.  The natural choice is to think of $S^5$ as
the surface $\sum_{I=1}^6(x^I)^2=1$ in $\IR^6$, and use the cartesian
coordinates.  We obtain the following metric:
\eqn\alchmetr{
ds_5^2(\al,\chi)\eql {a^2\over 2}{{\rm sech}\chi\over
\xi}(dx^IQ^{-1}_{IJ}
dx^J)+{a^2\over 2}{\sinh\chi\tanh\chi\over \xi^3} (x^IJ_{IJ}dx^J)^2\,,
}
where $Q$ is a diagonal matrix
with $Q_{11}=\ldots=Q_{44}=e^{-2\alpha}$ and $Q_{55}=Q_{66}=e^{4\alpha}$,
$J$ is an antisymmetric matrix with $J_{14}=J_{23}=J_{65}=1$,
$\xi^2\eql x^IQ_{IJ}x^J$.   The  warp factor is simply
\eqn\warp{
\Omega^2\eql \xi~\cosh\chi\,. }
The constant, $a$, has been introduced in \alchmetr\
to account for the arbitrary normalization of the Killing vectors.
The constant can be fixed by ensuring that \metrten\ and \alchmetr\
yield the proper metric for the round $S^5$ compactification
($\alpha =\chi =0$).  Specifically, one finds:
\eqn\anorm{a \eql \sqrt{2}~L_0 \ ,}
where $L_0$ is the radius of $AdS_5$ at the maximally supersymmetric
point.

The metric \alchmetr\ is valid for any configuration of scalar fields,
$\alpha$ and $\chi$, including ones that depend on space-time
coordinates.  It is invariant under $SU(2)\times U(1) \times U(1)$:
The tensor $Q$ is manifestly $SU(2)\times SU(2) \times U(1)$
invariant, while $J$ is invariant under $SU(3) \times U(1)$, with the
last $U(1)$ generated by $J$ itself.

We now pass to coordinates that are adapted to this
isometry. Introduce the complex coordinates based on the complex
structure $J$:
\eqn\complexcoordinates{
u^1\eql x^1+ix^4\,,\qquad u^2\eql x^2+i x^3\,,\qquad u^3\eql x^5-ix^6\,,
}
Now use the group action to reparamerize these coordinates, \ie\ let:
\eqn\angcoords{
\left(\matrix{u^1\cr u^2\cr}\right)\eql
 e^{-i\phi/2}\,\cos\theta\, g(\al_1,\al_2,\al_3)\,
\left(\matrix{1\cr 0\cr}\right)\,,\qquad
u^3\eql  e^{-i\phi}\,\sin\theta\,,}
where $g(\al_1,\al_2,\al_3)$ is an $SU(2)$ matrix expressed in terms of
your Euler angles.

Performing the change of variables and restricting to the critical point
we obtain
\eqn\themetric{\eqalign{
 ds_5^2&\eql {\sqrt{3}\over 8}a^2 (3-\cos(2\theta))^{1/2}\left(
 d\theta^2 +{\cos^2\theta\over
 3-\cos(2\theta)}((\sigma^1)^2+(\sigma^2)^2)
 +{\sin^2(2\theta)\over(3-\cos(2\theta))^2}(\sigma^3)^2  \right)\cr
 &\qquad + {\sqrt{3}\over 12}a^2(3-\cos(2\theta))^{1/2}\left(d\phi+
 {2\cos^2\theta\over 3-\cos(2\theta)} \sigma^3\right)^2\,,\cr}
}
where $\sigma^i$, $i=1,2,3$, are the $SU(2)$ invariant forms satisfying
$d\sigma^i=\sigma^j\wedge\sigma^k$, and
\eqn\wrnow{
\Omega^2\eql 2^{1/3}(1-{1\over 3}\cos(2\theta))^{1/2}\,.
}

The $SU(2)$ invariance is manifest, and the two $U(1)$'s correspond
to a $\phi$-translation and the rotation of $\sigma^1$ into
$\sigma^2$.  For future reference we will call these rotations
$U_\phi (1)$ and $U_\sigma(1)$ respectively.

We choose orthogonal frames $e^M=(e^m,e^a)$, $m=1,\ldots,5$,
$a=6,\ldots,10$, for the metric \metrten\ such that
\eqn\vielbeins{\eqalign{e^1&\propto dx^0\,,\qquad e^2 \propto
dx^1\,,\qquad e^3 \propto dx^2\,,\qquad
e^4 \propto dx^3\,, \qquad e^5 \propto dr\,,\cr
e^6&\propto d\theta\,,\qquad e^7\propto \sigma^1\,,\qquad e^8\propto
\sigma^2\,,\qquad e^9\propto \sigma^3\,,\qquad e^{10}\propto d\phi
+\ldots \ .  \cr}}
An explicit evaluation of the Ricci tensor yields the
following nonvanishing components
\eqn\adsric{\eqalign{
R_{11}&\eql-R_{22}\eql\ldots\eql-R_{55}\cr &
\eql {4\over L^2}\, \Omega(\theta)^{-2}
-{4\over \tilde a^2}\,\Omega(\theta)^{-10}\,
(\cos(4\theta)-10\cos(2\theta)+5)\,,\cr}
}
\eqn\ricone{
R_{66}\eql R_{99}\eql
-{8\over 3\,\tilde a^2}\,
\Omega(\theta)^{-10}\,
(9\,\cos(2\theta)-31)\,,}
\eqn\rictwo{
R_{77}\eql R_{88}\eql
{4\over 3\,\tilde a^2}\,\Omega(\theta)^{-10}\,
(\cos(4\theta)-18\cos(2\theta)+61)\,,
}
\eqn\ricthr{
R_{10\,10}\eql
{4\over 3\,\tilde a^2}\,\Omega(\theta)^{-10}\,
(5\cos(4\theta)-54\cos(2\theta)+117)\,,
}
and
\eqn\ricfor{
R_{9\,10}\eql
{4\over \tilde a^2}\,\sqrt{{2\over 3}}\,\Omega(\theta)^{-10}\,
(\sin(4\theta)-6\sin(2\theta))\,,
}
where
\eqn\atilde{
\tilde a^2\eql {27\over 4\,2^{2/3}}\,a^2\,.}

\newsec{The Ansatz and the solution}

The bosonic field equations of the IIB supergravity,
with the dilaton and axion set to zero read \JSIIB:
\eqn\tenein{
R_{MN}={1\over 6} F^{PQRS}{}_MF_{PQRSN}+
        {1\over 8}(G^{PQ}{}_MG^*_{PQN}+G^{*PQ}{}_MG_{PQN}-
        {1\over 6}g_{MN} G^{PQR}G^*_{PQR})\,,
}
\eqn\tenself{
F_{MNPQR}={1\over 5\,!}\,e\,\epsilon_{MNPQRSTUVW}F^{STUVW}\,,
}
\eqn\tenmaxwell{
\nabla^P G_{MNP}=-{2\over 3}\,i\,F_{MNPQR} G^{PQR}\,,
}
and
\eqn\tengsq{
G^{PQR}G_{PQR}\eql 0\,,
}
together with Bianchi identities
\eqn\Fbian{
dF\eql {1\over 8}\,{\rm Im}\, G\wedge G^*\,,
}
and
\eqn\Gbian{
dG\eql 0\,.
}

We are seeking a solution in which the 5-form field, $F$, is invariant
under the symmetries of $AdS_5$.  This leads to a generalized
Freund-Rubin Ansatz  \refs{\DiLithium,\PvNW}  that is consistent
with the warped product form of the metric  and satisfies the
self-dual field equations:
\eqn\fiveten{
F\eql -m \Omega^{-5}(e^1\wedge\ldots\wedge e^5+
e^6\wedge\ldots\wedge e^{10})\,.
}

To make the Ansatz for the potential, $B$, of the 3-index tensor, $G$,
one starts by considering $G$ at the linearized level.  The relevant
harmonics lie in the ${\bf 10}$ and ${\bf \overline{10}}$ of $SO(6)$,
and in the cartesian coordinates these can be represented as self-dual
and anti-self-dual combinations of $dx^I \wedge dx^J \wedge dx^K$.  We
want the $SU(3)$ invariant combination $du^1\wedge du^2\wedge du^3$
(and its complex conjugate).  Writing this in terms of the coordinates
\angcoords\ one finds the natural combination of the left invariant
$1$-forms: $\sigma^1 - i \sigma^2$, which transforms with charge $-1$
under $U_\sigma(1)$.  We also find that $G_{{\rm linear}} = du^1\wedge
du^2\wedge du^3$ depends upon $\phi$ as an overall phase of $e^{-2 i
\phi}$.  The $U(1)$ invariance of $G_{{\rm linear}}$ is thus a
combination
the $U_\sigma(1)$ with parameter $\beta$ and $U_\phi(1)$ with
parameter $-\beta/2$.  This must also be an invariance of the full $G$
tensor.

Recall now that in the IIB theory, the $G$ and $G^*$ tensors form an
$SL(2,\IR)$ doublet, with $G$ having charge $+1$ under $U(1) \in
SL(2,\IR)$.  Using this $U(1)$ we must be able to promote
$U_\sigma(1)$ and $U_\phi(1)$ to full invariances of $G$.  Put another
way the complete $G$ tensor must transform with the same phases as
$G_{{\rm linear}}$ under $U_\sigma(1)$ and $U_\phi(1)$.  This means
that the complete $G$ must depend upon $\phi$ as an overall phase
$e^{-2 i\phi}$, and must involve $\sigma^1$ and $\sigma^2$ only in the
combination: $\sigma^1 - i \sigma^2$.

This leads to the following Ansatz for the potential, $B$, of the
3-index tensor:
\eqn\Banz{
B\eql e^{-2i\phi}\left(a_1(\theta)d\theta\wedge(\sigma^1-i\sigma^2)+
a_2(\theta)(\sigma^1-i\sigma^2)\wedge\sigma^3\right)\,. }
It should also be noted that the chiral combination
$\sigma^1-i\sigma^2$ ensures \tengsq.  This is essential since
$G^2$ is the source of the dilaton, and this is constant in the five
dimensional solution.

The stress tensor for the $5$-form alone is
\eqn\fiveem{
T^{(5)}_{mn}\eql 4m^2\Omega^{-10} g_{mn}\,,\qquad
T^{(5)}_{ab}\eql -4m^2\Omega^{-10}g_{ab}\,,
}
where in our conventions ({\it cf.} \ \vielbeins) $m,n$ represent
flat indices along $AdS_5$, while $a,b$ are flat indices along the
internal manifold.

The self-duality of $F$ implies that $F_{M_1\ldots M_5}F^{M_1\ldots
M_5}=0$ so that the only contribution to the scalar curvature comes
from the 3-index tensor,
\eqn\scalcur{
R\eql {1\over 24} |G|^2\,. }
Substituting this back into the Einstein equations \tenein\ along the
$AdS_5$
directions determines $m$ (in \fiveem) and fixes the $AdS_5$ radius,
$L$, in terms of $a=\sqrt{2} L_0$:
\eqn\constants{
m\eql {4 \over 9 } { 2^{1/3}\over L_0}\,,\qquad
L\eql {3\over 2 \, 2^{2/3}} L_0\,.}
This generates the proper value of the cosmological
constant for the $AdS_5$ (cf.\ \cosmorat):
\eqn\anocom{
{\Lambda\over \Lambda_0}\eql \left({L\over L_0}\right)^2 \eql
{4 \, 2^{4/3} \over 9} \,.
}

By examining the Einstein equations \tenein\ along the internal space
one can eliminate $a_1(\theta)$ in terms of
$a_2(\theta)$. Substituting this into the Maxwell equations \tenmaxwell\
yields one second order and two first order differential equations for
$a_2(\theta)$, and these can be integrated. The result is
\eqn\solution{
a_1(\theta)\eql -{2\,2^{1/3}\over 9}\,i\,\,L^2 \cos\theta\,,\qquad
a_2(\theta)\eql - i\,{\sin(2\theta)\over
3-\cos(2\theta)}\,a_1(\theta)\,. }

Our result for the $B$-field is particularly
simple when written in terms of frames:
\eqn\Btens{
B \eql -{i \over \sqrt{3}}~e^{-2i\phi} ~
(e^6 + i e^9) \wedge(e^7 -i e^8)   \,. }

With this $B$-field all the equations of motion are satisfied,
and we have a solution that has the correct global symmetry and
cosmological constant.  It remains
to verify that the solution is indeed supersymmetric.

\newsec{Supersymmetry}

To find the unbroken supersymmetry we must solve
the equations $\delta \psi_\mu =0$; $\delta
\lambda =0$ in the IIB supergravity theory.
These equations have the form (for vanishing dilaton
and axion) \JSIIB:
\eqn\gravsusy{
D_M\epsilon+{i\over 480}\,F_{PQRST} \ga^{PQRST}\ga_M\epsilon
+ {1\over 96}G_{PQR}\left(\ga_{M}{}^{PQR}-
9\delta_M{}^P\ga^{QR}\right)\epsilon^* \eql 0\,,}
and
\eqn\halfsusy{
\ga^{MNP}G_{MNP}\epsilon \eql 0\,.}
We adopt exactly the same conventions as \JSIIB, except
that our frame indices in \vielbeins\ run from $1,\dots,10$ as opposed
to $0,\dots,9$.

We start by solving \halfsusy.  From \Btens\ it follows
that:
\eqn\formofG{G\eql dB\eql \omega\wedge B\,, }
where
\eqn\formofom{
\omega\eql -{3 i\over L \Omega} \,\left( e^{10} ~-~i\,
\sqrt{{2\over 3}}\, {\sin(2\theta) \over
(3-\cos(2\theta)) } \,e^{6} \right)\,. }
One can check that when contracted into $\gamma$-matrices
the differential form $\omega \wedge (e^6 + i e^9)$ yields
an invertible operator (at least for $\sin(2\theta)\not=0$).
It follows that \halfsusy\ is satisfied if and only if
the supersymmetry generators satisfy:
\eqn\Gcond{
(\ga^7-i\ga^8)\epsilon\eql 0 \, .}
This selects a 16 (real) parameter subspace
of chiral spinors.

The $SU(2)\times U(1)$ invariance of the background
implies that   $\epsilon=\epsilon(x^\mu,r,\theta,\phi)$ does not
depend on the $SU(2) \times U(1)$ directions.  Moreover, the
supersymmetry generators are a doublet of $SL(2,\IR)$, and this
can be used to fix the $\phi$-dependence of $\epsilon$ just as we
did in the Ansatz for $G$.
Similarly, the superconformal invariance on the $AdS_5$ fixes
the dependence on $x^\mu,r$, $\mu =1,\dots,4$.
These facts can also be verified by direct computation using \gravsusy.

The solutions for $\epsilon$ are most easily determined by taking
carefully chosen linear combinations of the equations \gravsusy,
\ie:
\eqn\combine{
\alpha_M\gamma^M\delta\psi_M+\alpha_N \gamma^N\delta\psi_N\eql 0\,, }
where there is no summation over repeated indices.  One chooses
the constants $\alpha_M$ and $\alpha_N$ so that the terms with
the $5$-index tensor, $F$, cancel.  For the directions along the
brane \combine\ reduces to
\eqn\difalong{
\gamma^m  {\partial\over\partial x^m}\epsilon
\eql \gamma^n  {\partial\over\partial x^n}\epsilon\,,\qquad
m,n\eql1,\ldots,4\,,}
while with one parallel and one in the transverse $r$-direction
\combine\ becomes
\eqn\partran{
2 L \gamma^m{\partial\over\partial x^m}\epsilon\eql
e^{r/L} \gamma^5 \left( 2 L {\partial\over\partial r}-1
\right)\epsilon\qquad m=1,\ldots,4\,. }
The general solution to these equations is a linear combination of
\eqn\spindepone{
\epsilon^{(+)} \eql e^{r/(2L)}\xi^{(+)}\,,}
\eqn\spindeptwo{
\epsilon^{(-)}\eql
\left(e^{r/(2L)}{x^\mu\gamma_\mu\over L}+e^{-r/(2L)}\gamma^5\right)
\xi^{(-)}\,, }
where $\xi^{(\pm)}=\xi^{(\pm)}(\theta,\phi)$ are $SU(2)$ invariant
chiral spinors, $\gamma^{11}\xi^{(\pm)}=\mp\xi^{(\pm)}$, that depend
on the internal coordinates only.

Using similar difference of variations along the brane and an $SU(2)$
direction, and simplifying the derivative terms using \spindepone\
or \spindeptwo\ and the independence on the $SU(2)$ coordinates,
yields an algebraic constraint:
\eqn\sigdep{
\left(\sqrt{2}\cos\theta~(\gamma^5-i\gamma^{10})-
\sqrt{3}\sin\theta~(\gamma^6+i\gamma^9)\right)\xi^{(\pm)}\eql 0\,.}
The two obvious solutions to this constraint are given by constant
spinors $\eta^{(+)}$ and $\eta^{(-)}$ which satisfy
\eqn\aniihop{
(\ga^5-i\ga^{10})\eta^{(\pm)}\eql 0\,,\qquad
(\ga^6+i\ga^{9})\eta^{(\pm)}\eql 0\,,\qquad
(\ga^7-i\ga^{8})\eta^{(\pm)}\eql 0\,,\qquad }
where, for completeness, we have included \Gcond.
It follows from this and the helicity of $\epsilon$ (\ie\
$\ga^{11}\epsilon = -\epsilon $) that:
\eqn\chiralbr{
\ga^1\ga^2\ga^3\ga^4\eta^{(\pm)}\eql \pm i\eta^{(\pm)}\,.}
The constraints \aniihop\ reduce the original $32$
supersymmetries by a factor of $8$ to four supersymmetries.
Indeed, the spinors, $\eta^{(\pm)}$, have opposite chirality along
the  brane and each is thus a single complex Weyl spinor in
four dimensions.

Associated to each solution, $\eta^{(\pm)}$, of the constraints
\aniihop\ is a  ``conjugate spinor'' that also solves \sigdep:
\eqn\nextsol{
\tilde \eta^{(\pm)}\eql (\sqrt{2} \cos\theta \,\gamma^5-\sqrt{3}
\sin\theta \,\gamma^6)\,\gamma^7 \,(\eta^{(\pm)})^*\,. }
Note that because of the complex conjugation, $\tilde \eta^{(\pm)}$
has opposite brane-helicity to $\eta^{(\pm)}$.

At this point we have the building blocks of the unbroken
supersymmetry.  One takes a linear combinations of $\eta^{(\pm)}$
and $\tilde \eta^{(\pm)}$ with coefficients that are arbitrary
functions of $\theta$ and $\phi$ and substitutes them into the
remaining supersymmetry variations.  The result is a
system of algebraic and first-order differential
equations.   The solution is
\eqn\possol{
\xi^{(\pm)}(\theta,\phi)\eql (3-\cos(2 \theta))^{-3/8}\,\left(
e^{-2 i\phi} \,\sin\theta \, \eta^{(\pm)} -
i(\sqrt{2}\cos\theta \,\gamma^5-\sqrt{3}
\sin\theta \,\gamma^6)\,\gamma^7 \,(\eta^{(\pm)})^*\right)\,.}
Remember that $\eta^{(\pm)}$ are independent, complex
Weyl spinors on the brane.  Thus $\epsilon^{(\pm)}$ provide an
$\cN =2$ supersymmetry in the bulk, while $\epsilon^{(+)}$, which
is independent of the brane coordinates (see \spindepone),
represents the $\cN =1$ supersymmetry on the brane.

Before concluding we wish to make two remarks about the form of these
supersymmetry generators.  First, when seeking supersymmetries in
complicated backgrounds it is often easier to start by finding the
brane supersymmetry generators \spindepone.  Having found these,
the other bulk supersymmetry in $AdS_5$, \spindeptwo, can be generated
using the superconformal symmetry.  For example, the conformal boost
between the time and $r$ directions corresponds to the Killing vector:
\eqn\Kvec{\eqalign{ K^1 \eql & {1 \over 2L} (\sum_{a=1}^4 (x^a)^2 +
L^2 e^{-2r/L}) \,,   \qquad K^5 \eql  -x^1 \,, \cr
\quad K^i \eql & {1 \over L}  x^1 x^i \,, \quad
i=2,3,4. }}
The infinitessimal change of coordinates: $x^\mu \to x^\mu +
\varepsilon  K^\mu$ induces a boost in the vielbein frame.
When this change of coordinates and local frame boost acts
on the spinor $\epsilon^{(+)}$ in \spindepone\ the result
is the spinor $\epsilon^{(-)}$ in \spindeptwo\ with
$\xi^{(-)} = \ga^1 \xi^{(+)}$.

The second issue has to do with dimensional reduction.  When
one reduces the ten-dimensional IIB action to five dimensions
one must perform rescalings of almost all the fields by
powers of the determinant of the metric on the internal
dimensions.  This is to make sure that the $\sqrt{g}$ factors
for the internal dimensions are not present in the five-dimensional
kinetic terms.  This is the origin of the warp factors.
In particular, if $\hat \psi_u$ is the five-dimensional
spin-$3/2$ field then for the metric \metrten\
one has  $\hat \psi_u = \Omega^{-1/2}   \psi_u$, and hence
the supersymmetry generators must be similarly related
$\hat \epsilon= \Omega^{-1/2} \epsilon$.  This extra factor
converts \possol\ into something a little more canonical.

\newsec{Final comments}

As we indicated in the introduction, the solution we have
exhibited here is primarily of significance because of the
of its relationship to a non-trivial supersymmetric fixed point
of large N, $\cN=4$ Yang-Mills.  It is also significant in that
it represents a highly non-trivial confirmation of consistent
truncation:  Our ten-dimensional solution has precisely the
global symmetries, supersymmetries, cosmological constant, and indeed
internal metric that were predicted from the five-dimensional
gauged $\cN=8$ supergravity \KPW.

The expression for the complete ten-dimensional metric as a
function of all the five-dimensional supergravity scalars was
first given in \KPW, and indeed, shortly after writing \FGPWa,
we computed the metric \metrten, and examined it in detail along the
flow and at the fixed point.   Similar results were recently
re-derived in \CLP.  However, the highly non-trivial task in finding the
solution presented here, and indeed in the general consistent
truncation story, is to handle the tensor field backgrounds and
supersymmetry generators beyond the linearized level,
and then solve both the equations of motion and the
equations $\delta \psi_\mu =0; \delta \lambda =0$.  Our expression,
\Btens, for the $B$-field is, in fact, rather simple when expressed in
terms of frames.  This feature almost certainly is a result of
the high level of symmetry in our solution.  However, the simplicity
of our result gives hope that more general solutions might be
attainable, and indeed the construction of the ten-dimensional
version of the complete flow of \FGPWa\ should be within reach.
Work on this is continuing.

\bigskip
\leftline{\bf Acknowledgements}

This work was supported in part
by funds provided by the DOE under grant number DE-FG03-84ER-40168.

\listrefs
\vfill
\eject
\end